\documentclass[conference]{IEEEtran}
\usepackage{graphicx}
\usepackage{url}

\begin{document}

\title{Identifying Coherent Anomalies in Multi-Scale Spatio-Temporal Data using Markov Random Fields}

\author{\IEEEauthorblockN{Adway Mitra}
\IEEEauthorblockA{International Center for Theoretical Sciences, Bangalore, India\\
Email: adway.cse@gmail.com}}

\maketitle

\begin{abstract}
Many physical processes involve spatio-temporal observations, which can be studied at different spatial and temporal scales. For example, rainfall data measured daily by rain gauges can be considered at daily, monthly or annual temporal scales, and local, grid-wise, region-wise or country-wise spatial scales. In this work, we focus on detection of anomalies in such multi-scale spatio-temporal data. We consider an anomaly as an event where the measured values over a spatio-temporally extended region are significantly different from their long-term means. However we aim to avoid setting any thresholds on the measured values and spatio-temporal sizes, because not only are thresholds subjective but also the long-term mean values often vary spatially and temporally. For this purpose we use spatio-Temporal Markov Random Field, where latent states indicate anomaly type (positive anomaly, negative anomaly, no anomaly/normal). Spatio-temporal coherence is maintained through suitable edge potentials. The model is extended to multiple spatio-temporal scales to achieve our second goal: anomalies at any scale should be defined both on the data at that scale, and also on anomalies at other scales. This allows us to trace an anomaly at a coarse scale to finer scales. For example, whether rainfall in a particular year is anomalous over a region should depend not only on the total volume of rainfall over the entire region, but also on whether there were such anomalies at the grid-scale, and the monthly scale. We use this approach to study rainfall anomalies over India -extremely diverse with respect to rainfall- for the period 1901-2011, and show its benefits over existing approaches.
\end{abstract}

\IEEEpeerreviewmaketitle

\section{Introduction}
The rapid progress in Data Science and sensing techniques in recent years has resulted in many physical processes being studied in a data-centric way. Most of such processes are spatio-temporal, where data measurements are carried out at multiple locations, at regular intervals of time. Dealing with spatio-temporal data is different from standard forms of data, because of spatio-temporal correlations and diversities. Measurements of the same quantity, say temperature, at two different places can have totally different statistical properties. But at locations that are close-by, the measurements are often strongly correlated. Again, the measurements may vary across time periodically, and there can be long-term trends also. For example, in most places, temperature at daytime is higher than at night (periodic oscillation), but as a result of global climate change, there is also a general increasing trend over time. Such complexities necessitate research specifically aimed at addressing problems regarding spatio-temporal data.

In this work we focus on a specific domain: rainfall over India. This phenomena has been a hot topic of research in the Climate Science and Hydrology community for several decades. Not only does it spawn many intriguing scientific questions, but it has tremendous social impact as a large number of people in India are dependent on the rains for their livelihood. Most places in India receive nearly $80\%$ of their annual rainfall from the South Asian Monsoon, which is active every year in the period June-September. But some places, mainly in the south-eastern parts of the peninsula, receive most of their rains in October-November, while some other places in the southern parts receive more rain in May. Some places in the north-eastern parts and along the western coast receive very heavy rainfall on over $60\%$ of the days each season, while places in the north-western region receive good rainfall only on a few days. The total volume of rainfall varies moderately across years, and even the spatial distribution of rainfall over the landmass, and monthly distribution over the season vary significantly from one year to another. The vagaries of the monsoon rainfall are discussed in detail in~\cite{monsoon}. 

Indian Meteorological Department (IMD) declares every year as normal or \emph{extreme} (exess rainfall or deficient rainfall), depending on the total volume of rainfall received over the entire country during the monsoon season. A year when the aggregate annual rainfall deviates from the long-term mean by at least one standard deviation is considered as extreme by them. However, it is also possible that in a year of deficient rainfall overall, some locations have normal or even excessive rainfall, and vice versa. Generally, in excessive-rainfall years, most places in the country receive more rainfall than their long-term means (which varies greatly across the country), but only a few locations will have extreme rainfall, which can be considered as the cause of the aggregate rainfall being extreme. Similarly, in such years, only a few months (usually in the monsoon season) receive excessive rainfall, while other months have only slightly above-average rain. Identifying coherent regions which received extreme rainfall in a year when countrywide annual rainfall was extreme (positive/negative) can help understand the phenomenon better, and also make suitable policies to deal with such events in future. It is also similarly important to identify stretches in time (perhaps a few months), in which the extreme rainfall was concentrated.
On the other hand, a year in which a large number of locations receive extreme rainfall in a few months should be considered as an extreme year even if  the aggregate annual rainfall is not off from the mean by over a standard deviation. This is because, spatio-temporally local extremes have a significant bearing on the society. Deficient rain can result in very poor yield of agricultural output, but excessive rains can cause floods along which much loss of life and property. So, our aim in this work is to discover spatio-temporal extremes in Indian rainfall at multiple spatial scales (country-wise and region-wise) and temporal scales (annual and monthly).
This is essentially a task of multi-scale spatio-temporal anomaly detection. We want these anomalies to be defined with respect to local spatial and temporal statistics of the locations instead of some global standard, due to tremendous diversity of rainfall characteristics, both spatially and temporally, across the dataset.

Anomaly Detection is a well-studied problem under Data Mining, and in recent years it has been studied well for spatio-temporal data. We discuss the relevant literature in Section 2. But our work is fundamentally different and novel from the existing literature for the following reasons: 1) We consider multi-scale data, and the anomalies at different scales are inter-dependent; 2) Within each scale, the anomalies are not limited to individual spatio-temporal locations, but to coherent zones, whose limits or sizes are not fixed. Also, within a zone constituting the anomaly, the measurement at every individual node need not be anomalous, as long as it is surrounded by anomalous locations and its indivdiual measurement is not far from the same type of anomaly. Basically a trade-off is made between spatio-temporal coherence and local measurements.

\section{Related Work on Anomaly Detection}

Anomaly Detection is well-studied area of Data Mining~\cite{anomalysurvey}. However, its main challenge is that anomalies cannot be precisely defined, and are very subjective in nature, and most papers on anomaly detection solve a specific formulation of the problem. Much of the work on anomaly detection is about classifying each individual data-point as normal or anomalous, with respect to either its immediate neighbors or the entire dataset. It is more difficult when we deal with collections of data-points rather than individually. 

One early approach~\cite{imageMRF} to \emph{spatial anomaly detection} was using \emph{Markov Random Fields}~\cite{MRF}, where the aim was to segment an image into regions, and simultaneously identify each region as normal/anomaly, depending on if they were similar or different from the other regions. MRFs along with various sophistications (such as unknown number of hidden states) have been used frequently for image segmentation~\cite{hdpmrf,infHMRF}. More recently, the Markov Random Field technique was used on worldwide gridded rainfall data, to identify major droughts of the world~\cite{MRFdrought}. Very recently,~\cite{GMRFanomaly} used Mixture of Gaussian Markov Random Fields to handle mutiple normal modes of their system, and provide variable-specific anomaly scores. \cite{SText} attempts to model temporal patterns of extreme values in spatio-temporal networks, taking into account spatial correlations on the extreme value parameters through an MRF-based prior. In this work, we continue to use the Markov Random Field approach, because it is a natural way to model spatio-temporal data, using nodes to represent spatio-temporal locations and edges to represent neighborhood.

Spatio-temporal anomaly detection for various applications have also explored other approaches. For example,~\cite{STDBN} and~\cite{STBNgas} both use Bayesian networks, in which they build process models for their applications on sensor networks related to environmental monitoring. The paper~\cite{STBNgas} discusses multiple types of anomalies where either individual sensor nodes, or multiple nodes at different parts of their network can get affected.  Other recent works for spatial or temporal anomaly detection include~\cite{localanom} (which attempts to localize anomalize spatially or temporally using graph-based, K-nearest-neighbor approach),~\cite{STdev}(which attempts to discover anomalies in spatio-temporal network data using large deviation theory) and~\cite{TempBilevel} (which attempts to discover similar but rare temporal patterns within time-series).

\section{Notations and Problem Definition}

\begin{figure}
\centering
\includegraphics[width=3.5in,height=1.5in]{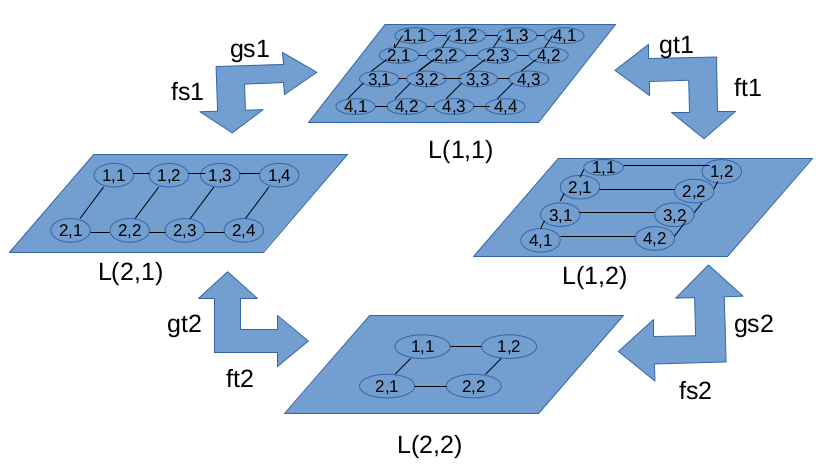}
\caption{Spatio-temporal locations at multiple scales, here shown for two spatial and two temporal scales. Each plane stands for a scale, and each circle for a spatio-temporal location. Spatial and temporal neighborhoods between the locations are indicated by the vertical and horizontal edges respectively. The thick edges indicate relations across the scales.}
\end{figure}

Consider a spatial-temporal dataset covering $S$ grid-locations, for which we have measurements of a physical variable (like rainfall) for $T$ time-points. For example, each location can be a grid cell of size $10Km-by-10Km$, and the frequency of measurement can be an hour. Now, these spatial and temporal intervals can be regrouped into coarser resolutions.  For example, these cells can be partitioned into larger cells of $25Km-by-25Km$, then into $50Km-by-50Km$ an so on. Temporally, the hourly intervals may be coalesced into six-hourly, daily, weekly, monthly, annual intervals. In principle, there are infinitely many ways in which such coarsening or downscaling of resolutions can be done, and these need not even be uniform. But without loss of generality, we assume that there are $L_{sp}$ spatial scales and $L_{tp}$ temporal scales to be considered for the task at hand.

Consider any particular spatio-temporal scale $(l,m)$ in this hierarchy, where $l$ denotes the spatial scale and $m$ denotes the temporal scale, i.e. $l\in \{0,L_{sp}\}$ and $m\in \{0,L_{tp}\}$. At this scale, denote by $S^l$ the number of spatial locations, and $T^m$ the number of time-points. Also, at each scale, each location $s$ has a set of neighboring locations $NB^{l}(s)$ in the same scale. For each pair of coordinates $(s,t)$ where $s\in \{1,S^l\}$ and $t\in \{1,T^m\}$ we have measurements of a variable, such as rainfall, denoted by $X^{l,m}(s,t)$.  The lowest scale, i.e. $(l=0,m=0)$ corresponds to the original dataset, with $S^0=S,T^0=T$ and $X^{0,0}$ are the same as the spatio-temporal measurements in the dataset. At higher scales, $X$ is obtained by averaging values from the lower scales. 

Denote by $Z^{l,m}(s,t)$ as the discrete ``state variable" at location $s$ and time $t$, which can take 2 or 3 values (depending on the application and scale), specifying if the location is having a positive or negative anomaly, or normal value, at that time. Unlike $X$, $Z$ is unknown and needs to be estimated. 

Every location in any spatial scale, say $l$, can be mapped to another location at the next higher (i.e. coarser) scale $(l+1)$, and also a set of locations at the next lower (finer) scale $(l-1)$. Let this mappings be called $f^l_{sp}$ and $g^l_{sp}$ respectively. Similarly, the function $f^m_{tp}$ can map time-points at scale $m$ to those at coarser scale $(m+1)$, and $g^m_{tp}$ can map them to finer scale $(m-1)$. The whole set-up is shown in Figure 1, where we consider 2 spatial and 2 temporal scales (i.e. $L_s=2,L_t=2$).

\subsection{Multi-Scale Spatio-Temporal Markov Random Field}
In our models, we will consider both the above variables $Z$ and $X$ as random variables, though $Z$ is latent and $X$ is observed. Markov Random Fields over a set of random variables is defined using a graph, where each node corresponds to one random variable, and on each edge/clique of this graph a potential function is defined -a function of the variables associated with the vertices covered by the edge/clique. The likelihood is defined as the product of all these functions.

For each of the spatio-temporal scales, say $(l,m)$, we consider a graph, where each node is associated with a spatio-temporal location. We have $S^l*T^m$ nodes for each of the $\{X^{lm}\}$ and $\{Z^{lm}\}$ variables. Next we define the edges at that scale, making use of the spatial neighborhoods $NB^l$ of locations at that scale. $Z$-nodes that are associated with adjacent spatial locations but same temporal location (eg. $Z^{lm}(s,t)$ and $Z^{lm}(s',t)$, where $s$ and $s'$ are adjacent) are connected by \emph{spatial edges}. Similarly, $Z$-nodes of each type that are associated with adjacent time-points but same spatial location (eg. $Z^{lm}(s,t)$ and $Z^{lm}(s,t+1)$) are connected by \emph{temporal edges}. Also, each $X^{lm}(s,t)$ is connected with $Z^{lm}(s,t)$ at the same spatio-temporal location, by \emph{data edges}.

Finally, we consider \emph{scale edges}, that connects the $Z$-variables at each spatio-temporal location at a scale, to the respective spatio-temporal graphs at the higher and lower scales. So, $Z^{lm}(s,t)$ is linked to $Z^{l+1,m}(f^l_{sp}(s),t)$ and 
$Z^{l,m+1}(s,f^m_{tp}(t))$ through scale edges. Figure 2 shows the model for two spatial scales, $(l=0,m=0)$ and $(l=1,m=0)$, where $S^1=2$. Figure 3 shows two temporal scales $(l=0,m=0)$, and $(l=0,m=1)$ where $T^1=T^0/2$. We illustrate the spatial, temporal, scale and data edges.
\begin{figure}
\centering
\includegraphics[width=3.5in,height=2.0in]{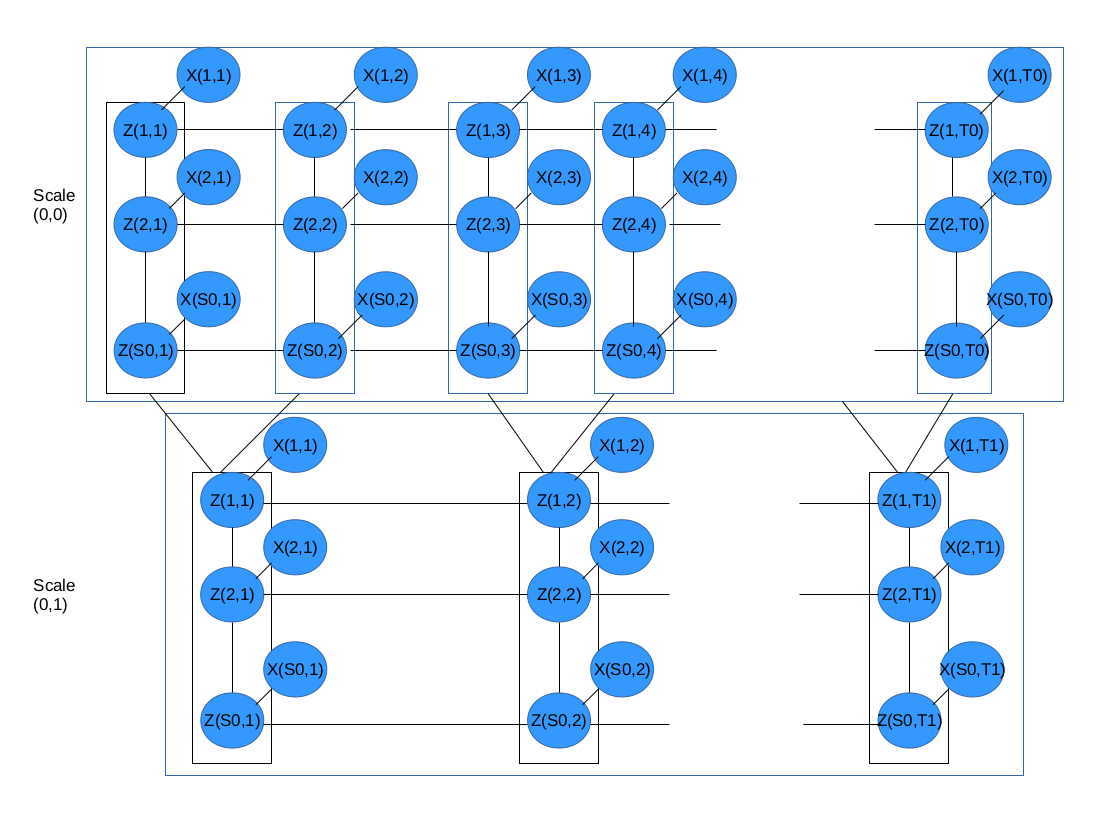}
\caption{Proposed MS-ST-MRF model for two temporal scales. Each node denotes a variable, specific to a spatio-temporal location. The horizontal and vertical axes indicate temporal and spatial dimensions respectively, and the edges parallel to them are the temporal and spatial edges respectively. The diagonal edges are scale edges, and short ones are data edges. The scale $m=1$ has half resolution than scale $m=0$.}
\end{figure}
\begin{figure}
\centering
\includegraphics[width=3.5in,height=1.6in]{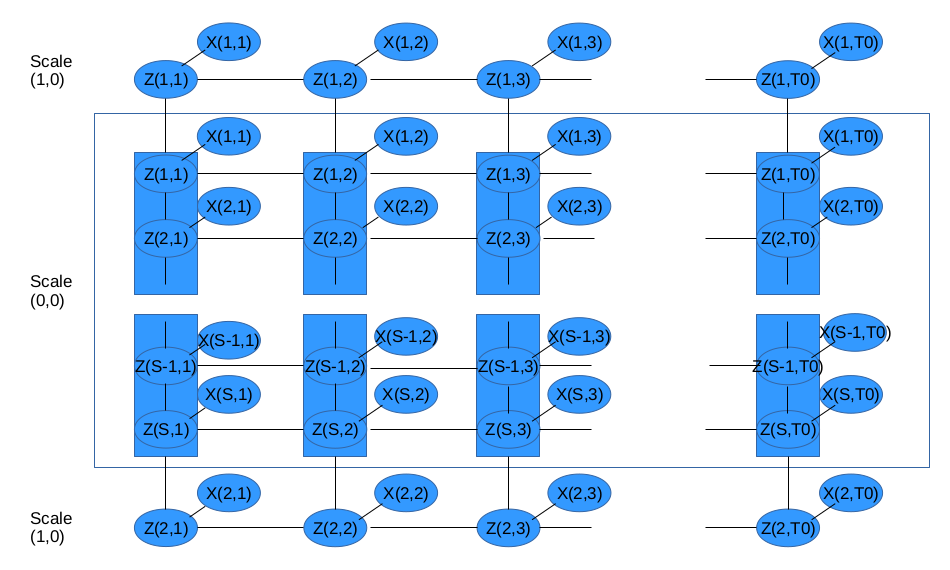}
\caption{Proposed MS-ST-MRF model for two spatial scales. Conventions same as Figure 2. The spatial scale $l=1$ contains just two location, each corresponding to half the spatial locations at scale $l=0$.}
\end{figure}

\subsection{Spatio-temporal Coherence}
As we have already discussed, our definition of anomalies is over spatio-temporally extended regions, not isolated points. So, one spatio-temporal location can be part of an anomaly only if some of its neighbors are also part of that same anomaly. In other words, the $Z$-variables are required to be \emph{spatio-temporally coherent}. We achieve this by \emph{MRF potential functions} on the spatial, temporal, scale and data edges.

In case of spatial and temporal edges, the purpose is to promote spatial and temporal coherence respectively. So, their potential functions are defined such that they take a high value if their end-points take the same value, and a low value otherwise. Mathematically,

\begin{footnotesize}
\begin{eqnarray}
\Psi^S(e) &= HIGH_{e} \textnormal{ if } Z^{lm}(s,t)=Z^{lm}(s',t); \\
&= LOW_{e} \textnormal{ otherwise } \nonumber \\
\Psi^T(f) &= HIGH_{f} \textnormal{ if } Z^{lm}(s,t)=Z^{lm}(s,t+1); \nonumber \\
&= LOW_{f} \textnormal{ otherwise }  \nonumber 
\end{eqnarray}
\end{footnotesize}
where spatial edge $e$ connects $Z^{lm}(s,t)$ and $Z^{lm}(s',t)$ ($s$ and $s'$ are neighboring locations, i.e. $s' \in NB^l(s)$), temporal edge $f$ connects $Z^{lm}(s,t)$ and $Z^{lm}(s,t+1)$ and each edge has two constant parameters - $HIGH$, $LOW$. 

Next we come to scale edges. As discussed in the introduction, we want to find corresponence between anomalies at one scale to those at its neighboring spatial and temporal scales. So, as with spatial or temporal edges, in case of scale edges also we want to define the potential functions which will be higher when both end-points are at the same anomaly state. However, every point at a  higher scale is associated with several points in the lower scale. The probability of a spatio-temporal location at a high scale being considered as anomolous should increase according to the number of locations under it at the lower scales that are considered having the same type of anomaly (in addition to its spatio-temporal neighbors and its own measurement). So, we define the scale edge potentials as follows:

\begin{footnotesize}
\begin{eqnarray}
\Psi^{SC}(g) &= exp(HIGH_g) \textnormal{ if } Z^{lm}(s,t)=Z^{l+1,m}(s',t); \\
&= exp(LOW_{e}) \textnormal{ otherwise } \nonumber \\
\Psi^{TC}(h) &= exp(HIGH_h) \textnormal{ if } Z^{lm}(s,t)=Z^{l,m+1}(s,t'); \nonumber \\
&= exp(LOW_{e}) \textnormal{ otherwise } \nonumber 
\end{eqnarray}
\end{footnotesize}
where $s'=f^l_{sp}(s)$ and $t'=f^m_{tp}(t)$.
The exponential term is used to count the number of matches in state values between location $(s,t)$ in scale $(l,m)$ and the locations under them in scales $(l-1,m)$ and $(l,m-1)$, when product is taken over all the edge potential functions to compute the full likelihood. This number plays a role in the likelihood of the state assignments. $Z^{lm}(s,t)$ is more likely to be 1 if a higher number of $Z$-variables coming under it in scales $(l-1,m)$ and $(l,m-1)$ are also in state 1.

In case of data edges, say $(Z^{lm}(s,t),X^{lm}(s,t))$, the potential function $\Psi$ is defined as the PDF of some suitable continuous distribution at $X^{lm}_{st}$ and the parameters of the distribution are determined by $Z^{lm}(s,t)$. For example if we use Gaussian distribution for $X$, then we will have a set of Gaussian parameters, one for each value in the discrete support-space of $Z$, and at any location $(s,t)$ we will use the Gaussian parameters corresponding to the value assigned to $Z^{lm}(s,t)$. The exact distribution to be used is dependent on the application, and it may be different at different scales.

\subsection{Inference}
The full likelihood of the model, for a particular assignment of all the variables, is given by the product of all the edge potentials. It can be written as:
\begin{footnotesize}
\begin{eqnarray}
\mathcal{L}(Z,X) = \prod_{l,m}^{L_{sp},L_{tp}}\prod_{s,t}^{S,T}\Psi^D(Z^{lm}(s,t),X^{lm}(s,t)) \\
\times \prod_{l,m}^{L_{sp},L_{tp}}\prod_{s,t}^{S,T-1}\Psi^T(Z^{lm}(s,t),Z^{lm}(s,t+1))  \nonumber \\
\times \prod_{l,m}^{L_{sp},L_{tp}}\prod_{s,t}^{S,T}\Psi^S(Z^{lm}(s,t),Z^{lm}(s',t))  \nonumber \\
\prod_{l,m}^{L_{sp},L_{tp}}\prod_{s,t}^{S,T}\Psi^{SC}(Z^{lm}(s,t),Z^{l+1,m}(f^l_{sp}(s),t))   \nonumber \\
\prod_{l,m}^{L_{sp},L_{tp}}\prod_{s,t}^{S,T}\Psi^{TC}(Z^{lm}(s,t),Z^{l,m+1}(s,f^l_{tp}(t)))   \nonumber
\end{eqnarray}	
\end{footnotesize}
We have observations for $X$ at all scales and spatio-temporal locations, but we do not know the values of $Z$. So we look for the assignment that maximizes the likelihood. We use Gibbs Sampling to infer them. We could have used a fully Bayesian approach by putting prior distributions on the various parameters, but in this work we consider them as user-defined. However, the parameters associated with the data edges are estimated iteratively along with the inference algorithm.

A very important step in iterative algorithms, for quick convergence is a good initialization. For this purpose, we first ignore all the spatial, temporal and scale edges, and make an initial estimate of all the $Z^{l,m}(s,t)$ variables conditioned only on the corresponding observations $X^{l,m}(s,t)$, which is effectively a problem of learning mixture models. 

Next, the Gibbs Sampling proceeds by sampling one latent variable at a time, keeping all others constant. This is easy because the model's likelihood function factorizes according to the edges, and for sampling the variable at each vertex we only need to consider the factors corresponding to the edges attached to that vertex, as all other factors remain unchanged. The Gibbs Sampling equation for $Z^{lm}_{st}$ at any spatio-temporal location $(s,t)$ and scale $(l,m)$ is given by:

\begin{footnotesize}
\begin{eqnarray}
&prob(Z^{l,m}(s,t)=k) \propto \Psi^D(k,X^{l,m}(s,t)) \\
&\times\prod_{\delta\in \pm 1}\Psi^T(Z^{l,m}(s,t+\delta),k)\times\prod_{s'\in NB^l(s)}\Psi^{S}(Z^{l,m}(s,'t),k) \nonumber \\
&\times\Psi^{SC}(X^{l+1,m}(f^{lm}_{sp}(s),t),k)\times \prod_{s'\in g^{l-1,m}_{sp}(s)}\Psi^{SC}(k,X^{l-1,m}(s',t)) \nonumber \\
&\times\Psi^{TC}(X^{l,m+1}(s,f^{lm}_{tp}(t)),k)\times \prod_{t'\in g^{l,m-1}_{tp}(t)}\Psi^{TC}(k,X^{l,m-1}(s,t')) \nonumber
\end{eqnarray}
\end{footnotesize}

We repeat this sampling process for many iterations, and collect samples of $Z$-variables at regular intervals. From these samples, the optimal values of $Z$ are computed.

Once the optimal assignment of the $Z$-variables has been done, we connect spatio-temporally locations at each scale having same $Z$-value, to form spatio-temporally coherent regions, each of which is associated with a single value in the support-space of $Z$. Each coherent region is a \emph{connected component of the spatio-temporal graph}. So, if $Z=1$ corresponds to high values of the climatic variable in question, then a spatio-temporally coherent region associated with $Z$-value of 1 defines a ``positive anomaly". Similarly, we can have ``negative anomalies" as a spatio-temporally coherent region associated with $Z$-value of 2. These anomalies regions are computed separately at each spatio-temporal scale, though we expect cross-scale correspondance between anomaly regions.

\section{Experimental Study}
In this section, we present our experimental results. We first describe our dataset and our model settings. Then we evaluate our results based on several measures, and compare them against baselines.

\subsection{Dataset}
As already discussed, our motivation in this paper is detection of spatio-temporal anomalies of rainfall over India. We consider a dataset that contains daily rainfall data measured over 357 grid-points, each of size $100Km-by-100Km$ covering the geopolitical landmass of India, for the period 1901-2011. Since the daily scale is too fine-grained and has too much uncertainties for meaningful definition of spatio-temporally coherent anomalies, we average the data temporally to monthly, and then yearly scales. Similarly, we spatially average the data to all-India scale, for both monthly and annual time-scales. 

In our terminology, there are two spatial scales ($L_{sp}=2$), where $l=0$ stands for grid-scale, and $l=1$ for the country scale. The number of locations are $S^0=357$ and $S^1=1$. There are also three temporal scales- daily ($m=0$), monthly ($m=1$) and annual ($m=2$). At these scales, number of time-points are $T^0=365*111=40515$, $T^1=12*111=1332$ and $T^2=111$. We neglect the temporal scale $m=0$ (daily). So for our analysis we will use the following scales $(0,1)$ -monthly mean at grid-level, $(0,2)$ -annual mean at grid-level, $(1,1)$ -monthly mean at country-level and $(1,2)$ -annual mean at country-level. The neighbors $NB$ are defined according to latitude-longitude coordinates of the grid-points, subject to the geo-political boundaries of India.

\subsection{Model Settings and Baseline Approaches}
To use MS-ST-MRF model on this data, for each scale we specify the number of anomaly states, parameters, and the distribution for the data measurements. For the scale $(0,1)$, i.e. for grid-locations at monthly scale, we consider two states for $Z$, which roughly correspond to ``high rain" and ``low rain". For the data edge potentials at this scale, we use Gamma distribution, i.e. $\Psi^D(Z^{(0,1)}(s,t), X^{(0,1)}(s,t)) \propto Gamma(X^{(0,1)}(s,t); \alpha_{smk},\beta_{smk})$ where $k=Z^{(0,1)}(s,t)$ and $m \in \{1,12\}$ indicates which of the 12 months $t$ is. The shape and scale parameters are specific to locations as well as months to account for the spatial and seasonal diversity of rainfall. These choices are based on the distribution of the data at this scale. For the other 3 scales, i.e. grid-wise rain at annual-scale $(0,2)$ and country-wide rain at both monthly and annual scales $(1,1)$ and $(1,2)$, we use 3 states for $Z$, which correspond to ``positive anomaly", ``negative anomaly" and ``normal condition". For the data edge potentials at all these scales we use Gaussian distributions, i.e. $\Psi^D(Z^{(l,m)}(s,t), X^{(l,m)}(s,t)) \propto Gaussian(X^{(l,m)}(s,t); \mu_{smk},\sigma_{smk})$ . Once again, this choice is made according to the distribution of data at these scales. Unlike the distribution at $(0,1)$-scale which is asymmetric with heavy right-tail, the distribution at other scales are symmetric. 

The parameters $\{\alpha,\beta,\mu,\sigma\}$ etc are estimated iteratively during inference. But the high and low values of the spatial, temporal and spatial edge potentials are fixed. Their values are computed according to correlations of the data time-series at the pairs of locations they connect.

Clearly, our aim is to assign values to the $Z$-variables, which the proposed model achieves by inference based on Gibbs Sampling. In assigning these variables, the proposed model promotes spatio-temporal coherence. For comparison, we also discuss a simple threshold-based technique to discover anomalies. For each location, we first estimate the mean and standard deviation for each of the 12 months, i.e. $(\mu_{sm},\sigma_{sm})$. At the scale $(0,1)$, the measurements $X^{(0,1)}(s,t)$ at each location are compared with the mean for that location and month, if higher $Z^{(0,1)}(s,t)$ is set to 1, else to 2. At other scales, if $X^{(l,m)}(s,t) > \mu^{lm}_{sm}+\sigma^{lm}_{sm}$, $Z^{lm}(s,t)$ is set to 1, if $X^{(l,m)}(s,t) < \mu^{lm}_{sm}-\sigma^{lm}_{sm}$, $Z^{lm}(s,t)$ is set to 2, and otherwise it is set to 3 (normal).

Among existing spatio-temporal anomaly detection techniques, the one that comes closest to this work is the model of~\cite{MRFdrought}, where a MRF was used to detect droughts (negative rainfall anomalies), using edge potentials to enforce spatio-temporal coherence. This is similar to our model at a single spatio-temporal scale, though it uses 2 latent states (negative anomaly and normal) instead of 3 as in our case, and Gaussian data likelihood. We accordingly modified it as ST-MRF to compare against our proposed multi-scale model. For inference of this model we use the Gibbs Sampling-based approach described above instead of the LP approach of~\cite{MRFdrought}.

\subsection{Evaluation Criteria}

We first consider the anomalies at each of the scales separately. We evaluate the \emph{mean spatial coherence} by both methods, which is the mean fraction of neighbors of any spatial location (say $s$) that has the same value of $Z$ as $s$. This mean is computed across all spatial locations and all time-points in each scale. Similarly, we evaluate the \emph{mean temporal coherence} by both methods, which is the  mean relative frequency with which any spatial location takes the same value of $Z$ at successive time-points. 

Next, we evaluate the anomalies which were identified by forming connected components of spatio-temporal locations that have the same value of $Z$ (as mentioned in Section IIIC). We compare their \emph{mean spatio-temporal sizes} (mean number of spatio-temporal locations covered by any  anomaly), \emph{mean spatial size} (mean number of unique spatial locations covered by any anomaly) and \emph{mean temporal size} (mean number of unique time-points covered by any anomaly).

Next, we study the inter-scale interactions of the anomalies. We are interested in answering three questions: 1) In any year of positive (respectively negative) anomaly at all-India scale, which regions (spatially contiguous set of locations) had positive (respectively negative) annual anomaly? 3) In any month of positive anomaly at all-India scale, which regions had positive anomaly? 3) In any year of positive anomaly at all-India scale, which period of successive months had positive anomaly? To answer these questions we compute the \emph{mean number of months and grid-locations in state 1} (positive anomaly), i.e. number of months satisfying $Z^{1,1}(1,m)=1$ and number of locations satisfying $Z^{0,2}(s,y)=1$ in years of positive anomalies at all-India scale, i.e. years satisfying $Z^{1,2}(y)=1$. Similarly, for years satisfying $Z^{1,2}(y)=2$ we compute the \emph{mean number of months and grid-locations in state 2}, i.e. satisfying $Z^{1,1}(1,m)=2$ and $Z^{0,2}(s,y)=2$. Finally, we also compare the \emph{spatial and temporal coherence of these locations and months}, i.e. the mean fraction of neighboring grid-locations or months having the same value of $Z$. We compare the results obtained by the three techniques of estimating $Z$ (i.e. by MS-ST-MRF, single-scale ST-MRF (similar to~\cite{MRFdrought}) and the local threshold-based alternative discussed above).

Finally, in each of the above cases, we also evaluate the \emph{mean value of rainfall at the anomaly locations}. The unit we use is millimeters per day per location, for all scales. Clearly, when we aim to look for spatio-temporally coherent zones, we have to include some intermediate locations where the measurement is not extreme, and exclude some isolated locations where it is extreme. In the trade-off, we need to ensure that intensities of the anomalies are not significantly lost in pursuit of spatio-temporal properties.

\subsection{Results}
We show the results in Table I for single-scale anomalies and in Table II,III and IV for multi-scale anomalies. All the tables clearly show that MRF-based methods result in increased spatio-temporal coherence. In Table I we find that mean spatial, temporal and spatio-temporal sizes are higher in case of the MRF-based methods, especially for MS-ST-MRF, compared to the baseline, which is predictable. But surprisingly, even with respect to the intensity of the anomalies (quantified by mean rainfall) the MRF-based approaches, especially MS-ST-MRF, outperform the baseline as the positive anomalies have higher mean rainfall and the negative anomalies have lower. 

In Table II, we again find similar patterns, as the MRF-based methods have better temporal coherence and anomaly intensity than the baseline, and the proposed MS-ST-MRF dominates ST-MRF. The number of months assigned to anomaly states by the MRF methods is greater than that by the baseline. In contrast, tables III and IV show that more locations are assigned to anomaly states in case of the baseline compared to the MRF methods. This is understandable since MRF-based methods wipe out the isolated anomaly locations due to their coherence property. In Table II, the reverse effect was seen because the data distributions at that scale makes setting of thresholds difficult. Tables III and IV show that the proposed MS-ST-MRF achieves the best spatial coherence, though the anomaly intensities are not always the best. 

\begin{tiny}
\begin{table}
\begin{tabular}{| c || c | c || c | c || c | c || c | c |}
\hline
Approach & STS1 & STS2 & SS1 & SS2 & TS1 & TS2 &X1 &X2\\
\hline 
Threshold & 25.5 & 123.8 & 21.3 & 75.9 & 1.3 & 1.6 & 4.8 & 3.6\\
STMRF[6]    & \textbf{65.0} & 372.3 & \textbf{34.7} & 105 & \textbf{1.7} & \textbf{2.8} & \textbf{4.9} & 3.0\\
MSSTMRF& 63.8 & \textbf{372.9} & 34.4 & \textbf{106} & \textbf{1.7}     & \textbf{2.8} & \textbf{4.9} & \textbf{2.8}\\
\hline
\end{tabular}
\begin{tabular}{| c || c | c || c | c || c | c || c | c |}
\hline
Approach  & STS1 & STS2 & SS1 & SS2 & TS1 & TS2 &X1 &X2\\
\hline 
Threshold & 6.7 & 6.5 & 6.1 & 5.7 & 1.2 & 1.2 & 5.0 & 2.4\\
STMRF[6]     & 13.1 & 17.6 & 9.6 & 10.6 & 1.5 & 1.9 & \textbf{5.6} & 2.3\\
MSSTMRF & \textbf{15.8} & \textbf{20.4} & \textbf{11.8} & \textbf{12.9} & \textbf{1.8} & \textbf{1.9} & \textbf{5.6} & \textbf{2.2}\\
\hline
\end{tabular}
\caption{\tiny{We compare the mean spatio-temporal sizes (STS1,STS2), spatial sizes (SS1,SS2), temporal sizes (TS1,TS2) and mean measurements (X1,X2) for positive and negative anomalies at the scale of grid-points and months (upper table), and grid-points and years (lower table) }}
\end{table}
\end{tiny}

\begin{tiny}
\begin{table}
\begin{tabular}{| c || c | c | c || c | c | c |}
\hline
Approach & $\#Z1$ & tcoh1 & mn(X1) & $\#Z2$ & tcoh2 & mn(X2)\\
\hline 
Threshold & 3.4 & 0.33 & 5.4 & 3.8 & 0.35 & 3.5\\
STMRF[6]   &  4.7   & 0.34 & \textbf{5.5} & 7.6 & 0.5 & 3.4\\
MSSTMRF   & 4.5 & \textbf{0.38} & 5.2 & 7.1 & \textbf{0.61}& \textbf{3.3}\\
\hline
\end{tabular}
\caption{\tiny{We compare the mean number ($\#Z1$ and $\#Z2$) of months in positive and negative anomaly states in years of positive and negative anomalies respectively by the three approaches. We also compare the mean temporal coherence (tcoh1,tcoh2) and the mean rainfall $mn(X1)$, $mn(X2)$ for such months, for both anomaly types}}

\begin{tabular}{| c || c | c | c || c | c | c |}
\hline
Approach & $\#Z1$ & scoh1 & mn(X1) & $\#Z2$ & scoh2 & mn(X2)\\
\hline 
Threshold & 103 & 0.6 & 4.9 & 106 & 0.65 & \textbf{2.1} \\
STMRF[6]   & 68 & 0.62 & 4.9 & 51 & 0.7 & 2.2\\
MSSTMRF   & 79 & \textbf{0.66} & \textbf{5.1} & 80  & \textbf{0.72} & 2.3 \\
\hline
\end{tabular}
\caption{\tiny{We compare the mean number ($\#Z1$ and $\#Z2$) of locations in positive and negative anomaly states in years of positive and negative anomalies respectively by the three approaches. We also compare the mean spatial coherence (scoh1,scoh2) and the mean rainfall $mn(X1)$, $mn(X2)$ for such locations, for both anomaly types}}

\begin{tabular}{| c || c | c | c || c | c | c |}
\hline
Approach  & $\#Z1$ & scoh1 & mn(X1) & $\#Z2$ & scoh2 & mn(X2)\\
\hline 
Threshold  & 207 & 0.83 & 6.5 & 305 & 0.93 & \textbf{1.7}\\
STMRF[6]      & 163 & 0.84 & \textbf{12} & 248 & 0.92 & 2.2\\
MSSTMRF & 198 & \textbf{0.87} & 6.2 & 285 & \textbf{0.94} & 1.9\\
\hline
\end{tabular}
\caption{\tiny{We compare the mean number ($\#Z1$ and $\#Z2$) of locations in positive and negative anomaly states in months of positive and negative anomalies respectively by the three approaches. We also compare the mean spatial coherence (scoh1,scoh2) and the mean rainfall $mn(X1)$, $mn(X2)$ for such locations, for both anomaly types}}
\end{table}
\end{tiny}


\section{Conclusions}
In this paper, we approached a novel aspect of the spatio-temporal anomaly detection problem: when the anomalies are defined not for individual data-points but for spatio-temporally coherent regions of flexible size and extents, and where the data is multi-scale, with anomalies at different scales related to each other. The work was motivated by a very important practical problem: identification of anomalies in Indian rainfall. The solution proposed is fairly general, and may be applicable to other domains also.

\end{document}